\documentclass[runningheads]{cl2emult}

\usepackage{makeidx}  % allows index generation
\usepackage{graphicx} % standard LaTeX graphics tool
\usepackage{subeqnar} % subnumbers individual equations
\usepackage{multicol} % used for the two-column index
\usepackage{cropmark} % cropmarks for pages without
\usepackage{eso}      % placeholder for figures
\makeindex            % used for the subject index
\def\gsim{\stackrel{>}{\sim}}

\def\yrs{{\rm \ years}}

\def\etal{{\em et al.~}}
\def\Ms{M$_{\odot}$}

\def\HH{H$_2$ }
\def\H2{H$_2$ }
\def\HHp{H$_2^+$ }

\def\Hm{H$^-$ }

\begin{document}
\title*{First Structure Formation and the First Stars}
\toctitle{First Structure Formation and the First Stars}
% allows explicit linebreak for the table of content
%
%
\titlerunning{First Structure Formation}
% allows abbreviation of title, if the full title is too long
% to fit in the running head
%
\author{Michael L. Norman\inst{1}
\and Tom Abel\inst{2}
\and Greg Bryan\inst{3}}
\authorrunning{Michael Norman et al.}
% if there are more than two authors,
% please abbreviate author list for running head
%
%
\institute{University of Illinois, Urbana, IL 61801
\and Harvard University, Cambridge, MA 02138
\and MIT, Cambridge, MA 02138}

\maketitle              % typesets the title of the contribution

\begin{abstract}
We discuss the results of recent 3D simulations of first structure formation in
relationship to the formation of the first stars. On the basis of a new,
high-resolution AMR simulation (spatial dynamic range = $3 \times 10^7$), 
we conclude that the first stars are likely to be massive.
\end{abstract}

\section{First Cosmological Objects}

%success of hierarchical models
Hierarchical theories of structure formation such as Cold Dark Matter (CDM) and 
its variants have been very successful in accounting for the existence of 
galaxies, galaxy clusters and cosmological large scale structure in the low 
redshift universe (cf.  Ostriker 1993, Bertschinger 1997). The cynic 
would say that this is what they were invented to do.  
However, the models also make definite predictions at earlier epochs and 
on different mass scales which can be checked observationally.
Some proof that these models are {\it essentially correct} is the remarkable 
agreement achieved between the observed and predicted properties of the 
Lyman alpha forest at $2 < z < 4$ (e.g., Zhang et al. 1997, Rauch 
1998)---something the models were not designed to do. Encouraged by 
this agreement, we may ask: {\it what were the first objects to form in 
such models?} 

%first cosmological objects
Assuming the CDM power spectrum extends to very small mass scales,
the first cosmological objects to form are small dark matter halos 
collapsing from small scale density fluctuations at high redshift 
($z \sim 30$) (Couchman \& Rees 1986, Tegmark \etal 1997). 
Halos whose mass is significantly
less than the cosmological Jeans mass $M \sim 10^{4-5} M_{\odot} $
would not be able
to trap the baryonic fluid, and hence not form astrophysical objects. 
Conversely, halos of approximately this mass or greater would be expected 
to form highly condensed objects (stars, black holes, etc.)
provided the gas can cool and transport angular momentum. 
At the temperatures and densities prevailing in such halos,
the primordial gas cools inefficiently by collisional excitation
of \HH molecules which exist in low abundance. As described in more detail
below, \HH forms via nonequilibrium gas phase reactions in which the 
tiny post-recombination electron abundance acts as a catalyst. 

This chemistry, combined with the nonlinear dynamics of halo formation and
mergers, makes this epoch of structure formation exceedingly complex
and interesting. We have termed this epoch {\it first 
structure formation}, and have explored it numerically in recent papers 
(Abel et al. 1998a,b; Abel, Bryan \& Norman 1998; 2000; Norman, Abel 
\& Bryan 1999). Here we review our numerical methodologies and findings,
and discuss their implications to the first stars.

\section{Relevance to First Stars}

%relevance to first stars
First structure formation is relevant to the theme of this conference as it is 
likely that the first stars are formed as a byproduct. Note this is not a 
theoretical certainty since it has been variously argued that the first 
objects are not stars in the usual sense, but rather objects at both extremes 
of the mass spectrum, ranging from supermassive black holes (Silk 1998) 
to Jupiter sized ``clumpuscules" (Combes \& Pfenniger 1998). Since the 
mass function of the first objects is not known, much of the theoretical 
literature consists of making various assumptions about the primordial 
mass function (PMF) and working out the cosmological consequences.  
Carr, Bond \& Arnett (1984) presented a comprehensive 
review of the prevailing ideas as of 1984, many of which are still relevant 
today. Couchman \& Rees (1986) placed the matter in a modern
 cosmological framework.
Both papers highlighted the impact of massive stars and possibly very 
massive objects (VMOs) on heating, ionizing and enriching the 
pregalactic medium with heavy elements. 

A few authors have attempted 
to compute the PMF from first principles: Silk (1983) argued for a flat 
PMF between $0.1-100 M_{\odot}$ on the basis of a linear analysis of 
thermal instability in a primordial gas near the critical density for $H_2$ 
formation via three body reactions. Padoan, Jiminez \& Jones (1997) 
applied the statistical model for star formation of Padoan (1995) to 
primordial globular cluster formation, and found a PMF slightly 
shallower than the Miller-Scalo (1979) IMF with a lower cutoff of $0.2 
M_{\odot}$. 

Given the theoretical uncertainties, let us pose the
following question: {\it do 
stars form in the first nonlinear structures in hierarchical models?} 
Indications from our recent numerical simulations is that they do.
In this paper, we follow our historical developments which lead us toward
this conclusion.

\section{\H2 Formation and the Minimum Mass to Cool}

\begin{figure}
\centering
\includegraphics[width=0.8\textwidth]{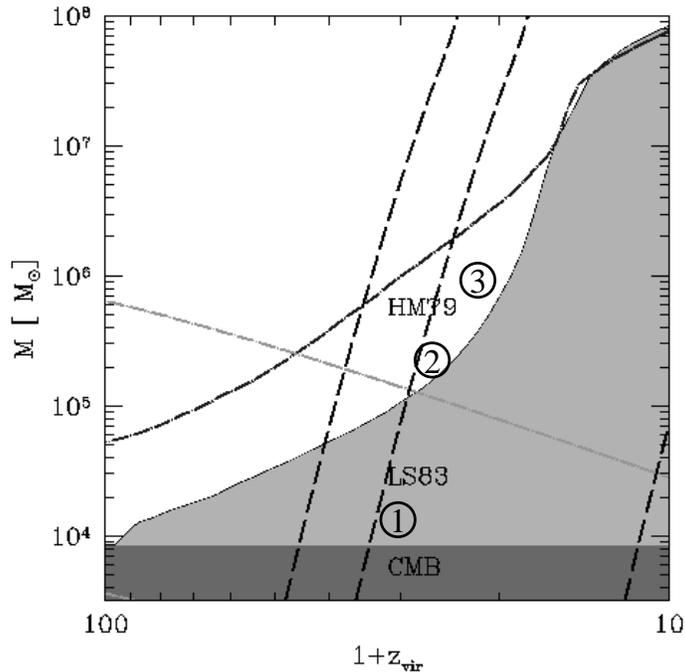}
\caption{The horizontal dark shaded region shows the mass scale for which 
the virial temperature equals the CMB temperature. The light shaded
area (labeled LS83) represents the domain of parameters for which collapsed
structures cannot cool. This curve is computed assuming a spherical collapse
model as in Tegmark \etal (1997), except we use the Lepp \& Shull (1983)
$H_2$ cooling function. Only above the light shaded area are structures
believed to be able to collapse and cool via \H2. The three heavy dashed 
lines plot the collapse redshift given by linear theory for the standard
CDM spectrum for $1\sigma, 3\sigma$, and $4\sigma$ perturbations. The
dot-dashed line represents the Jeans mass at 18$\pi^2$ times the background
density. The heavy dot-dashed line labeled HM79 is the minimum mass line
computed by Tegmark \etal based on the Hollenbach \& McKee (1979) cooling
function. The circled numbers denote the evolution of the halo discussed
in the text.} 
\label{minimass}
\end{figure}

After recombination, the universe is filled with a neutral, metal-free
gas of H and He and trace amounts of D and Li. 
How can such a gas cool to form stars?
Long ago it was realized that small amounts of molecular hydrogen will form
in a primordial gas via two reaction pathways:
the \Hm channel (McDowell 1961):
\begin{subeqnarray}
H + e^- \rightarrow H^- + h\nu, \label{f1a}\\
H^+ + H \rightarrow H_2 + e^- . \label{f1b}
\end{subeqnarray}
and the $H_2^+$ channel (Saslaw \& Zipoy 1967):
\begin{subeqnarray}
H + H^+ \rightarrow H_2^+ + h\nu, \label{f2a}\\
H_2^+ + H \rightarrow H_2 + H^+ . \label{f2b}
\end{subeqnarray}
At high redshifts $z > 100$ the fragile \Hm molecule is photodissociated 
by the CMB, and thus the \HHp channel dominates the production of \HH.
At lower redshifts, the situation is reversed and the \Hm channel
dominates. We see from Eq. (1) that free electrons catalyze 
the formation of \HH via the \Hm channel. Where do they come from? 
Well, the post-recombination primordial gas is not totally neutral. 
This is because the recombination time is longer
than the cooling time in such a diffuse gas, and the electron abundance is out
of equilibrium at a given temperature. Detailed nonequilibrium calculations
show that there is a residual ionization fraction of about $x_e \sim 10^{-4}$
at $z \sim 100$ (Galli \& Palla 1998). This is enough to form \HH with an 
abundance of $f_{H_2}\equiv n_{H_2}/n_H \sim 10^{-6}$. 

As gas concentrates in the potential wells of dark matter halos, the 
\HH fraction increases according to a simple analytic result (Abel \etal 1998a).
In the absence of an external
UV background at gas temperatures below 6000 K, one can integrate the
rate equations for the free electron and \H2 fractions in a collapsing
gas cloud to get:
\begin{eqnarray}
f_{H_2}(t)&=&f_{H_2}(t=0)+\frac{k_{PA}}{k_{rec}} ln(x_0 n_H k_{rec} t+1) \nonumber \\
          &=&f_{H_2}(t=0)+1.0\times 10^{-8} T_{vir}^{1.53} ln (t/t_{rec}+1),
\end{eqnarray}
where $k_{PA}, k_{rec}, t_{rec}, x_0$, and $n_H$ are the rate coefficients
of photo-attachment of \Hm and recombination to neutral hydrogen, the 
initial recombination timescale, ionized fraction, and neutral hydrogen
number density, respectively. Given the \H2 cooling function, one
can ask at what concentration will the cooling time be less than the Hubble
time for a spherical top hat perturbation of mass M collapsing at redshift
z? This analysis was first carried out by Abel (1995) and later by Tegmark
\etal (1997), who found a rather universal result: $f_{H_2}(crit) \sim 5
\times 10^{-4}$. Fig. (\ref{minimass}) shows the locus of critical 
cloud masses as a function of redshift for two choices of the
cooling function, which is still somewhat uncertain (for a recent
discussion of the uncertainties, see Tin\'{e}, Lepp \& Dalgarno 1998).
As can bee seen, extremely rare $4\sigma$ peaks of mass 
$\gsim 5 \times 10^4 M_{\odot}$ will collapse and cool at $z \sim 40$,
whereas more typical $2.5\sigma$ fluctuations of mass 
$\gsim 5 \times 10^5 M_{\odot}$ will collapse and cool at $z \sim 25$. 
The evolution of such a peak described in detail below is shown by the circled 
numbers in Fig. 1.

\section{Simulating First Structure Formation}

%do stars form in the first structures?
First structure formation
is a well-posed problem both physically and computationally, since for a 
given cosmogony the thermodynamic properties, baryon content, and 
chemical composition are specified for the entire universe at high 
redshifts, at least in a statistical sense. Futhermore, the dominant physical 
laws are readily identified: the general theory of relativity describing the 
evolution of the background spacetime geometry and particle geodesics, 
the Euler equations governing the motion of the baryonic fluid in an 
expanding universe, and the primordial kinetic rate equations determining 
the chemical processes. Radiation backgrounds, save the CMB, are 
entirely absent prior to first structure formation in standard models. 
The complete set of equations governing first structure formation in
a cosmological framework are given in Anninos \etal (1997).

%well-posed initial conditions
Unlike traditional star formation calculations, which are plagued by a lack 
of knowledge of appropriate initial conditions, the initial conditions for 
first structure formation calculations are well defined---namely, 
a Gaussian random field of density fluctuations characterized by a power 
spectrum whose shape can be computed theoretically. 
The amplitude of the fluctuations is constrained by 
observations on a variety of scales and redshifts (Ostriker 1993, Croft et 
al. 1999, Nusser \& Haehnelt 1999). In the calculations described here, 
we assume a cluster normalized standard cold dark matter (SCDM) model 
with the following parameters: $H_0=50$ km/s, $\Omega_0=1, 
\Omega_b=0.06, \sigma_8=0.7$. 

As in all star formation calculations, resolving the relevant length scales 
is the principal technical challenge. A simple estimate of what is required 
is to compare the baryonic Jeans length at a typical collapse redshift of 25 
to the solar radius:
\begin{equation}
\frac{\lambda_J}{ r_{\odot }}=5.7 pc (1+z)^{-1/2} / r_{\odot} \equiv 5 
\times 10^7
\end{equation}
Adding a couple more decades in scale to model the cosmological 
environment brings the total spatial dynamic range to around $10^{10}$. 
This is easily achievable in 1D spherical symmetry with logarithmic or 
adaptive grids. However, the all-important 
effects of hierarchical mergers, tidal forces, and fragmentation all demand 
3D simulations. Until recently, such dynamic ranges were impossible to
achieve in 3D.
 
We have developed two high-resolution,
3D numerical codes for solving the first structure formatiion equations set
forth in Anninos \etal (1997).
The first, more primitive code HERCULES (Anninos \etal 1994), uses fixed 
nested Eulerian grids for an effective resolution of $512^3$ in regions 
of interest. This code was used to carry out the first self-consistent
simulations of first structure formation (Abel \etal 1998) in which we
verified the predictions of Tegmark \etal (1997) and quantified the
cooled mass fraction in primordial halos. 
The second, more powerful code ENZO (Bryan \& Norman
1997, Norman \& Bryan 1999), utilizes an adaptive hierarchical mesh of
arbitrary depth to achieve unprecedented resolution wherever it is needed.

Using adaptive mesh refinement (AMR), 
we have previously reported the results of a 
simulation of first structure formation which achieved a spatial dynamic 
range of 262,144 (Abel, Bryan \& Norman 1998, 2000; Norman, Abel 
\& Bryan 1999). Here we update that result with a new simulation which 
achieves an additional factor of 128 in resolution for a total spatial 
dynamic range of $3 \times 10^7$. While not $10^{10}$, this is sufficient 
to characterize the properties of the collapsing protostellar cloud core. 
The calculation is done in a comoving box 128 kpc on a side. The starting 
redshift is z=100. At the stopping redshift z=19.1, the range of
resolved scales is $6.4 kpc \gsim \ell \gsim 2 \time 10^{-4} pc$. The mass
resolution in the initial conditions within the refined region are
0.53(8.96) \Ms in the gas (dark matter). The refinement criteria 
ensure that: (1) the local Jeans length is resolved by at least 4 grid
zones, and (2) that no cell contain more than 4 times the initial
mass element (0.53 \Ms).

\section{Core Formation}

\begin{figure}
\centering
\includegraphics[width=0.8\textwidth]{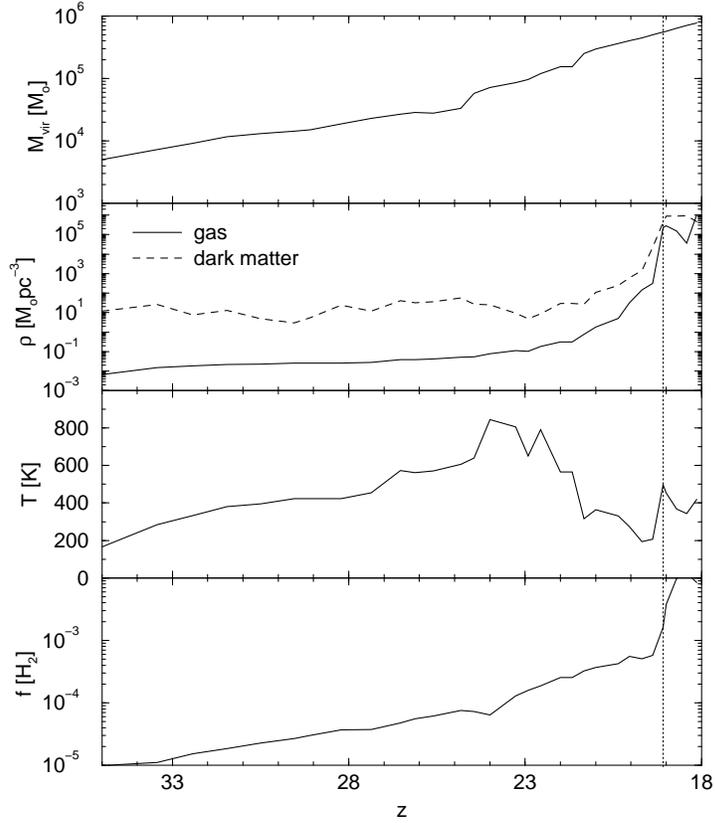}
\label{evolution}
\caption{\footnotesize The top panel shows the evolution of the virial
mass of the most massive clump as a function of redshift.  The remaining
panels show the density (both dark and baryonic), the temperature, and
the molecular hydrogen mass fraction at the central point of that clump.  The
central point is defined as the point with the highest baryon density.
Finite gas pressure prevents baryons from clumping as
much as the dark matter at redshifts $\gsim 23$.  Cooling begins
in earnest once the \H2 fraction reaches a few $\times 10^{-4}$,
lowering the central Jeans mass.  At $z \sim 19$ a dense core begins to
form and collapse with characteristic mass $\sim 200$ \Ms.}
\end{figure}

To illustrate the physical mechanisms at work during the formation of
the first cosmological object in our simulation, we show the evolution
of various quantities in Fig. 2.  The top panel of
this plot shows the virial mass of the largest object in the
simulation volume.  We divide the evolution up into four intervals.
In the first, before a redshift of about 33 (point 1 in Fig. 1), 
the Jeans mass in the
baryonic component is larger than the mass of any non-linear
perturbation.  Therefore, the only collapsed objects are dark-matter
dominated, and the baryonic field is quite smooth.  
%(We remind the
%reader that a change in the adopted cosmological model would modify
%the timing, but not the nature, of the collapse.)

In the second epoch, $23 < z < 33$, as the non-linear mass increases, 
gas becomes trapped in the gravitational potential wells of the
dark matter halos. However, this gas is not sufficiently dense to 
cool and the primordial entropy of the gas prevents dense cores from
forming.  This is shown in the second frame of Fig. 2
by a large gap between the central baryonic and dark matter densities
(note that while the central dark matter density is limited by resolution, the
baryonic is not, so the true difference is even larger).  As mergers
continue and the mass of the largest clump increases, its temperature
also grows, as shown in the third panel of this figure.  The \HH
fraction also increases (bottom panel).

By $z \sim 23$ (point 2 in Fig. 1), 
enough \HH has formed (a few $\times 10^{-4}$) 
that cooling begins to be
important.  During this third phase, the central temperature decreases
and the gas density increases.  However, the collapse is quasistatic
rather than a runaway freefall
because around this point in the evolution, the central
density reaches $n \sim 10^4$ cm$^{-2}$, and the excited states of \HH
are in LTE.  This results in a cooling time which is nearly
independent of density rather than in the low--density limit where
$t_{cool} \sim \rho^{-1}$ (e.g. Lepp \& Shull 1983).

Finally, at $z \sim 19$ (point 3 in Fig. 1), 
a very small dense core forms and reaches the
highest resolution that we allowed the code to produce.  We follow the
collapse to a final central density of $10^{12}$ cm$^{-3}$, well above
the critical density for the formation of \H2 via 3--body reactions
(see Palla \etal 1983). These reactions are included, as well as an
escape probability treatment of radiative transfer of the molecular
line radiation (Abel \etal 2000) which becomes important at these
densities. Finite word length and the lack of an appropriate equation
of state forced the termination of the calculation at this point.  

\section{Core Structure}

\begin{figure}
\centering
\includegraphics[width=0.8\textwidth]{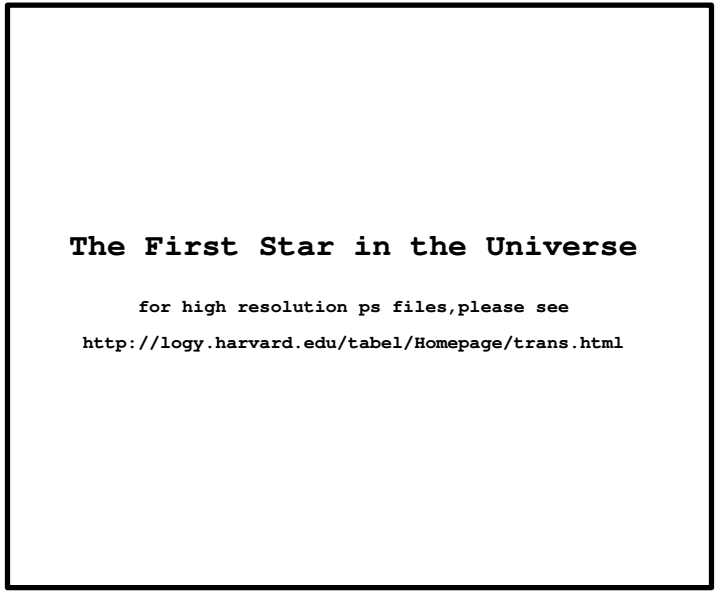}
\caption[]{Million-fold zoom showing the multiscale structure of
a primordial protostellar cloud core in the center
of a low mass halo at z=19.1. Plotted is the logarithm
of the baryon overdensity on a slice passing through
the densest structure on the grid. Zoom proceeds clockwise
from upper left. Linear scales are proper. The smallest grid
cell in the center of the cloud core in the last frame is
$2 \time 10^{-4}$ pc = 43 AU.} 
\label{first_star}
\end{figure}

The structure of the cloud at the end of the calculation is
shown in Fig. 4. The quantity plotted is the logarithm of the
baryon overdensity $\rho/<\rho>$ on a slice passing through the
densest cell at z=19.1. Proceeding from large scales to small
scales we see: (6 kpc)--the filamentary cosmological density field, with
a dense halo at the intersection of several filaments; (600 pc)--a virialized
halo with $M_{vir}\sim 4 \times 10^{5}$ \Ms ~and $r_{vir}\sim 100$ pc;
(60pc)--a dense baryonic core in the center of the cooling halo. From
scales of a few pc on down, the core is self-gravitationally bound and
collapsing. 

Despite the non-spherical structure of the cloud, mass-weighted
spherical averages of various quantities centered on the density maximum
 provide insight into the physical processes governing the formation and 
evolution of the core. A schematic based on a careful analysis of
these radial profiles is shown in Fig. 4.  

\begin{figure}
\centering
\includegraphics[width=0.8\textwidth]{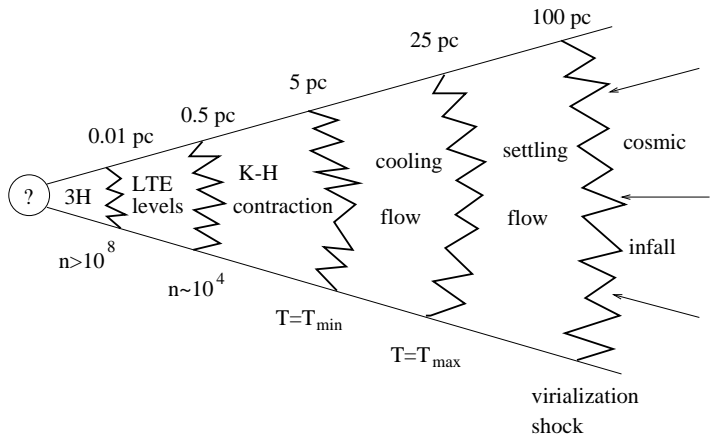}
\caption[]{Schematic of the structure of the primordial
protostellar cloud core based on spherical averages
around the density maximum at z=19.1. Jagged lines
demarkate different physical regions.}
\label{cartoon}
\end{figure}

Jagged lines in Fig. 4 indicate approximate radii where the
material makes a kinematic, thermodynamic, or other significant
physical transition.
Proceeding form large to small radii, we have the virialization
shock ($r_{vir} \approx 100$ pc); the cooling radius 
($r_{cool} \approx 25$ pc), defined where the
gas temperature reaches its maximum value $T_{vir}\sim 1000K$; 
the radius where the gas reaches the minimum temperature allowed
by \H2 cooling $\sim$ 200K
($r_{T_{min}} \approx 5$ pc); the radius where 
the \H2level population enters LTE ($r_{LTE} \approx .5$ pc) defined
where $n = 10^4$ cm$^{-3}$;
and the radius where 3--body \H2 formation reactions become
important ($r_{3H} \approx 10^{-2}$ pc), 
defined where  $n = 10^8$ cm$^{-3}$.

In the settling zone, the \HH number fraction rises from $7\times 10^{-6}$
to near $f_{H_2}(crit)$. Here \H2 cooling is negligible.
Inside $r_{T_{max}}$, a cooling flow is established reducing the 
gas temperature to $T_{min}$.
At $r_{T_{min}}$, the sound crossing time becomes
substantially shorter than the cooling time. This suggests that inside
this radius the core is contracting quasi--hydrostatically on the cooling
time scale, which approaches its constant high--density value at small
radii.   
This constant cooling time of $\sim 10^5\yrs$ sets the time scale of
the evolution of the fragment until it can turn fully molecular via
three body associations. Inside $r \sim 0.3$ pc, the enclosed baryonic 
mass of $\sim 200 M_{\odot}$ exceeds the local Bonnor-Ebert mass, 
implying this material is gravitationally unstable.
However, due to the inefficient cooling, its collapse is subsonic. 
The radius where $M > M_{BE}$ defines our protostellar cloud core.

\section{Fragmentation?}

Inside $r_{3H}$ atomic hydrogen is completely converted into \H2 via 
the 3--body reaction. This increases the cooling rate sufficiently
to initiate a freefall collapse in the central part of the core.  
What will be the fate of this collapsing cloud? 
Omukai \& Nishi (1998) have simulated the evolution of a collapsing,
spherically symmetric (1D) primordial cloud to stellar density including all
relevant physical processes. Coincidentally, their initial conditions
are very close to our final state. 
Our results agree with theirs where our solutions overlap.
Based on their results, we can say
that if the cloud does not fragment, a massive star will be formed.
Adding a small amount of angular momentum to the core does not change this
conclusion (Bate 1998), who found a dynamical bar mode instability
efficiently transports angular momentum outward in a 3D simulation
of galactic protostellar collapse.
A third possibility is that the cloud breaks
up into low mass stars via thermal instability in the quasi-hydrostatic
phase. Silk (1983) has argued that, due to the enhanced cooling from
the 3--body produced \HH, fragmentation of this core might ensue
and continue
until individual fragments are opacity limited (i.e. they become
opaque to their cooling radiation). We do not see any evidence of
this. Our code, which resolves the Jeans mass everywhere and at
all times, would not suppress the growth of these instabilities 
if they exist. 

Rather, inspection of the density distribution on the smallest
resolvable scales shows a centrally concentrated, object
of mass $\sim 10$ \Ms ~collapsing supersonically with $v_r \sim$
-4 km/s. The local free fall time is a few hundred years.
The protostar is embedded in a flattened nebula with a pronounced
one-arm spiral structure, suggesting an operative angular momentum
transport mechanism.

\section{Implications}

The direct implications of our results are that the first stars, here
defined to be those which form in a gas of zero metallicity, are massive. 
Here massive means of order the gravitationally unstable core which forms 
them $\sim 200$ \Ms. Where does this mass scale come from? 
We argue that it comes from the peculiar properties of a gas
cooling solely through \H2 line excitation, which sets a minimum temperature
for the gas of a few hundred K. The other factor is that above
$n_{LTE} \approx 10^{4}$ cm$^{-3}$, the cooling time becomes independent of
density, meaning that gravitationally bound clouds will evolve
quasi-statically until the onset of 3--body production of \H2
at much higher densities. 

The Bonnor-Ebert mass at $n=n_{LTE}$ and $T=T_{min}$ 
is 240 \Ms, close to our simulated core mass.  
We suggest that cold gas in the centers of primordial
low mass halos fragments into cores of this mass, each of which forms
a massive star (10-100 \Ms), or possibly a binary. These stars
would be strong sources of UV radiation prior to supernova and/or
black hole formation. Recent analyses (Haiman, Abel \&
Rees 1999; Omukai \& Nishi 1999) show that the UV
radiation from a single O star per halo will destroy \H2 both locally
and globally, quenching further star formation by this mechanism.
Due to the low binding energy of the parent halo, metals ejected by the 
supernova explosion would be returned to the pre-galactic medium.
Pockets of enriched gas collecting in the cores of more massive 
halos formed by subsequent mergers would form stars of various
metallicities in the familiar but complicated way we have been studying
for decades in Galactic molecular clouds. A key question, which
must await further numerical investigations, is whether there is
a pause in the star formation history of the universe, or rather
a smooth transition from the formation of the first stars to the
``next stars" as suggested by simulations of Ostriker \& Gnedin (1996).

%INDEX%%%%%%%%%%%%%%%%%%%%%%%%%%%%%%%%%%%%%%%%%%%%%%%%%%%%%%%%%%%%%%%
\clearpage
\addcontentsline{toc}{section}{Index}
\flushbottom
\printindex
%%%%%%%%%%%%%%%%%%%%%%%%%%%%%%%%%%%%%%%%%%%%%%%%%%%%%%%%%%%%%%%%%%%%%

\end{document}